\documentclass[onecolumn,showkeys,showpacs,superscriptaddress,notitlepage]{revtex4-1}
\usepackage{graphicx}
\usepackage{amsmath,amsfonts,amssymb}
\usepackage{morefloats}
\usepackage{color}
\usepackage{subcaption}

\newcommand{\noi}{\noindent}
\newcommand{\be}{\begin{equation}}
\newcommand{\ee}{\end{equation}}
\newcommand{\al}{\alpha}
\newcommand{\bt}{\beta}
\newcommand{\ab}{\alpha\beta}

\begin{document}
\title[ ]{Evolution of gossip-based indirect reciprocity on a bipartite network}

\author{Francesca Giardini}\email{f.giardini@rug.nl}
\affiliation{Faculty of Behavioral and Social Sciences, Department of Sociology, University of Groningen, Groningen (Netherlands)}
\affiliation{LABSS (Laboratory of Agent Based Social Simulation),
Institute of Cognitive Science and Technology,
National Research Council (CNR), 
Via Palestro 32, 00185 Rome, Italy}
\author{Daniele Vilone}\email{daniele.vilone@gmail.com}
\affiliation{LABSS (Laboratory of Agent Based Social Simulation),
Institute of Cognitive Science and Technology,
National Research Council (CNR), 
Via Palestro 32, 00185 Rome, Italy}
\affiliation{Grupo Interdisciplinar de Sistemas Complejos (GISC), Departamento de Matem\'aticas,
Universidad Carlos III de Madrid, 28911 Legan\'es (Spain)}

\maketitle

\ 

\

\

\section*{Abstract}
Cooperation can be supported by indirect reciprocity via reputation.Thanks to gossip, reputations are built and circulated and humans can identify defectors and ostracise them. However, the evolutionary stability of gossip is allegedly undermined by the fact that it is more error-prone that direct observation, whereas ostracism could be ineffective if the partner selection mechanism is not robust. The aim of this work is to investigate the conditions under which the combination of gossip and ostracism might support cooperation in groups of different sizes. We are also interested in exploring the extent to which errors in transmission might undermine the reliability of gossip as a mechanism for identifying defectors. Our results show that a large quantity of gossip is necessary to support cooperation, and that group structure can mitigate the effects of errors in transmission. 

\ 

\noi \textbf{Keywords:} {Gossip, Evolution of Cooperation, Bi-partite Graphs.} \\
\\


\section{Introduction}
Cooperation among individuals is essential for their survival, in human and animal societies. Human beings are an intrinsically social species and most of our evolutionary success can be attributed to our highly developed ability to cooperate  with each other. This ability is especially important in groups, where individuals need to coordinate their actions in order to achieve personal benefits that cannot be obtained without cooperation. However, those who do not contribute but reap the collective benefits are better off than cooperators~\cite{har68}.

In models of indirect reciprocity~\cite{now98a,now98b,vaq13}, cooperation can thrive when information about others is acquired either via direct observation, or via "image score", a reliable and publicly visible indication of one's past cooperative behaviour~\cite{oht06}.
When modelled as simple scores, pro-social reputations are evolutionary stable only if they track behaviour with the same accuracy as direct experience~\cite{rob08}. Cooperation becomes fragile when errors are possible, that is, when there is an even small probability for an individual to record a good partner as a bad one or vice-versa~\cite{uch13}. 

Image score is effective in supporting group cooperation, but only when group size of individuals playing a Public Goods Game (PGG) does not exceed four~\cite{suz05}. When group size increases, there is a concomitant decrease in the frequency of cooperation, showing that indirect reciprocity, even when supported by an image score mechanism, is not effective in large groups. The authors explain this decline in cooperation as due to the difficulty of observing reputations of many individuals in large communities. However, when agents are placed on a bipartite graph and they can actively select their group members, image score becomes effective in sustaining cooperation, even for groups of 20 individuals~\cite{vil14}. 

Notwithstanding its effectiveness in supporting cooperation in models of indirect reciprocity, image score is limited by its reliance on direct observation. Thanks to language, humans are able to overcome this limitation and can exchange information about each other, thus isolating defectors and selecting cooperative partners~\cite{bur08,gia12}.

Thanks to gossip, we can map our social group~\cite{dun04}, learn about its rules~\cite{fos04}, and enforce social norms~\cite{glu63}, among other things. Gossip is crucial to make information about known cheaters travel within the network, thus allowing for identification of defectors, and it has a strong influence on the behaviour of participants in an economic experiment, even when they can rely on direct observation of others' actions~\cite{som07}.

Gossip is also relatively more effective than punishment in promoting cooperation across a four-round PGG, it increases participants’ gains and also efficiency, whereas punishment significantly decreases participants’ earnings~\cite{wu16}. In a computational study, gossipers who could actively select their group members and avoid ill-reputed agents are able to outperform free-riders and punishers in groups of 25 agents, whereas in smaller groups the combination of gossiping and material punishment is more successful in increasing cooperation levels~\cite{vil14}.

However, the evolutionary stability of gossip in supporting cooperation has been questioned. According to Nowak and Sigmund~\cite{now05}, one of the main limitations of gossip consists in its being unreliable, while Ohtsuki, Iwasa and Nowak~\cite{oht09} assume that errors in observing interactions, and the resulting unreliable reputations, inevitably cause gossip-based indirect reciprocity to collapse.

The aim of this study is to investigate the impact of gossip quantity and quality on cooperation levels in a population of artificial agents playing a PGG on a bipartite network. In a bipartite graph the mesoscopic level of the interactions is better depicted~\cite{gar11,new93} than in a classical one-mode network, and it also makes group choices more relevant. As an illustration, two individuals belonging to the same three groups are ``more'' connected than two other individuals who share the membership of a single group, so that in this way it is possible to take into account also the quality and the weight of the connections (indeed, a similar result can be obtained by means of weighted networks, which however do not represent explicitly the group structure~\cite{new04} underlying the network).  Indeed, this kind of representation of the relations among individuals has been already shown to be more efficient when the interactions considered are competitive, as in the Prisoner's Dilemma Game or the PGG itself~\cite{szo09,pen12}.

Embedding agents into a bipartite graph allows us to investigate the relationship between gossip and ostracism. This form of punishment, defined as being ignored or excluded by another individual or group of individuals~\cite{wil07a}, is effective in promoting cooperative behaviours, favouring investments in the collective goods and maintaining social order~\cite{cin05,ouw05}. Maier-Rigaud and colleagues~\cite{mai05} show that in laboratory experiments, participants in a PGG with ostracism opportunities can increase contribution levels and, unlike monetary punishment, ostracism also has a significant positive effect on net earnings. 
In groups and small-scale societies, ostracism can result from being negatively gossiped about~\cite{ble76,ste04,glu63,glu68}, and a combination of these two mechanisms could effectively support cooperation. In a laboratory experiment, Feinberg and colleagues show that a combination of gossip and ostracism leads to restore the collective good at the end of the game, after an initial shrinking of group earnings~\cite{fei14}.

Given our interest for gossip in human societies, we also consider essential to investigate the relative effects of two ways of misreporting information about others. We do not distinguish here between malicious gossip, as intentionally misreported information, and random noise, and we focus only on the latter. However, we reckon that the direction of misreporting should be taken into account. A cooperator can be considered as a cheater, and then excluded from the interaction, or a defector could be erroneously included among cooperators. The former is an {\it exclusion error}, while the latter is an {\it inclusion error} and little is known about their respective effects on cooperation. 

We developed a model of gossip-based cooperation with the aim of addressing two main challenges related to the evolution of cooperation: 1. the impact of quantity and quality of gossip; and 2. the effect of network structure.
In order to test whether privately exchanged information could support the emergence of cooperation  we designed an agent-based model in which agents were embedded into groups composed by a fixed amount of players who could select a number of new players depending on direct experience and on reputational information acquired through gossip. The game consisted of a sequence of PGG rounds and gossip rounds, going from 1 to 5. The longer this sequence, the more gossip information agents could exchange in order to update other agents' reputations. We also introduced errors in transmission, in order to test whether and to what extent cooperators could survive when information about cheaters was not reliable. 
We measured how the levels of cooperation and the average reputation of the agents varied in response to the quantity and reliability of gossip, as well as in response to the severity of ostracism.

\ 

\section{Results}

\ 

We performed our simulations with different values of the model parameters $F$, $n$, $M$, $g_p$, $q_{\al}$, $q_{\bt}$ and $q_{\al\bt}$: these parameters will be precisely defined in the Methods section, anyway they are already summarized in Table~\ref{table1}. The results presented here are all averaged over $2000$ independent realizations.

\begin{table}[!ht]
\caption{
{\bf Summary of the model parameters.}}
  \centering
\begin{tabular}{|l|l|l|}
\hline
\multicolumn{1}{|c|}{\bf Variable} & \multicolumn{1}{|c|}{\bf Description} & \multicolumn{1}{|c|}{\bf Notes}\\ \hline
$M$ & number of groups & $\in \mbox{\bf N}$ \\ \hline
$n$ & size of groups & $\in \mbox{\bf N}$ \\ \hline
$F$ & fraction of fixed group members & $\in[0,1]$ \\ \hline
$L$ & total size of the system & $L=nFM$ \\ \hline
$q_{\al}$ & probability of excluding errors & $\in[0,1]$ \\ \hline
$q_{\bt}$ & probability of including errors & $\in[0,1]$ \\ \hline
$q_{\al\bt}$ & probability of bidirectional errors & $\in[0,1]$ \\ \hline
\end{tabular}
\label{table1}
\end{table}

 \ 
 
The variables of interest for understanding the effects of gossip and ostracism on cooperation are the average cooperator density $\rho$ ({\it i.e.}, the fraction of agents
adopting the strategy C in the population), and the average reputation $S$. The latter indicates the average value of the average score each agent has of the others over the entire population :

\be
S=\langle \sigma(i,j)\rangle_{i,j}=\frac{1}{L(L-1)}\sum_{i,j}\sigma(i,j) \ . 
\label{av_im}
\ee 

\noi In general, such quantities evolve in time ({\it i.e.}, through generations), so that we specify them as $\rho(t)$ and $S(t)$.
Also their final values $\rho^{\infty}$ and $S^{\infty}$, are taken into consideration in the description of the results.

\begin{figure}[h]
  \centering
\includegraphics[width=6.5cm]{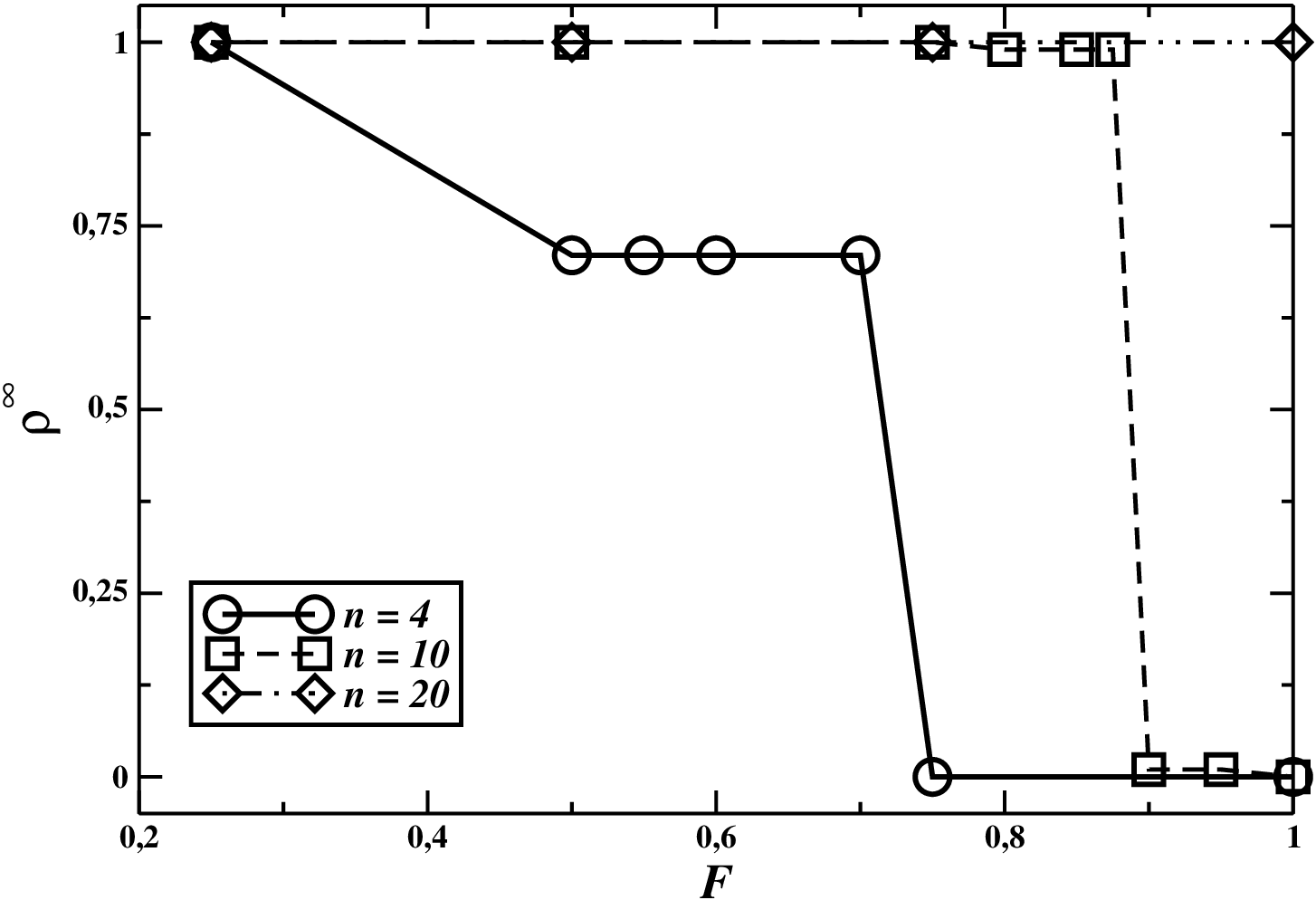}
\includegraphics[width=6.5cm]{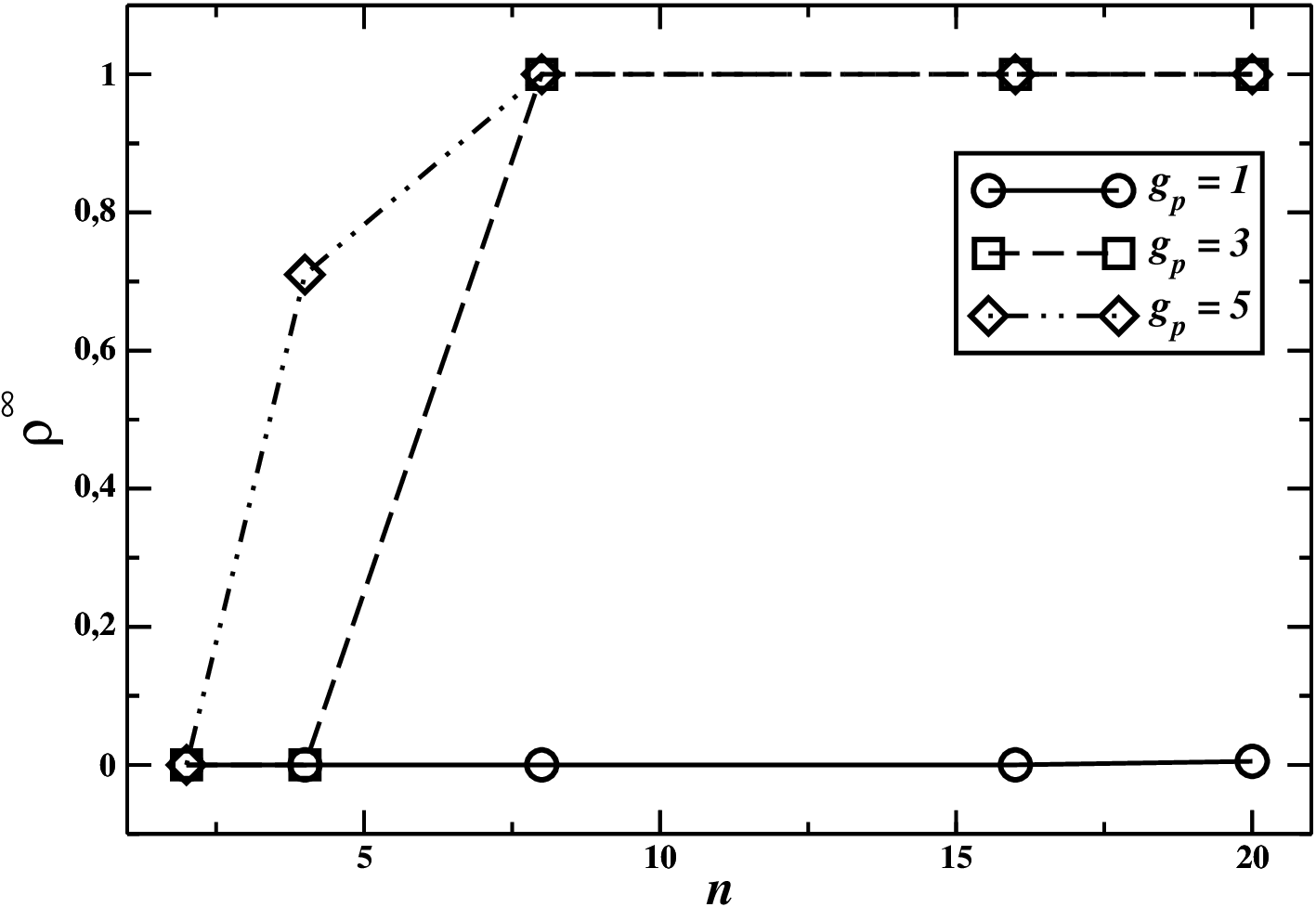}
\caption{
Left figure: final cooperator density as a function of the fraction $F$ of fixed members of the PGG groups for $g_p=5$, $n=4,\ 10$ and $20$, $M=100$ for $n=4$, $M=40$ for $n=10$, and $M=20$ for $n=20$; perfect information. Right figure: final cooperator density as a function of the group size $n$ for $L=200$, $F=0.5$, and $g_p=1,\ 3,\ 5$; perfect information (no errors). Please notice that we have $L=nFM$.}
\label{fin_coopF0.5_F}
\end{figure}

Figure~\ref{fin_coopF0.5_F} (left graph) shows the final density of cooperators as a function of the fraction of fixed members in PGG groups, $F$, for two different values of $n$ and $g_p=5$. In large groups, full cooperation is achieved when gossip is abundant ($g_p=5$). In smaller groups, the final cooperation level remains high for smaller values of $F$, but goes to zero when $F$ increases. 
When the opportunities for partner choice are limited, cooperation levels dramatically decrease as $F$ increases. Figure~\ref{fin_coopF0.5_F} (left graph) shows the importance of the partner selection mechanism: when $F$ becomes large enough to increase the number of fixed members in each PGG group then the final cooperation level goes down. For example, with $n=4$, $\rho^{\infty}$ undergoes a step-like reduction at $F=0.25$ (fixed members increase from 1 to 2), $F=0.5$ (fixed members from 2 to 3), until the final cooperation disappears for $F\geq0.75$, when all the four components of a group are permanent and there is no more partner selection. The same thing happens with different values of $n$. It is worth noticing that in the baseline version of the model, without partner choice ($F=1$) and a single gossip exchange ($g_p=1$), cooperative behaviours get rapidly extinct for both small and large group size after few generations, and the average reputation (as already stated, completely irrelevant for the evolution of the system) is negative.

Also the amount of  gossip exchanged plays a crucial role in supporting cooperation: Figure~\ref{fin_coopF0.5_F} (right graph) shows the behaviour of the final cooperator density as a function of $n$ for $L=200$, $F=0.5$, and three different values of $g_p$. 
Without gossip, cooperation cannot be sustained for any value of $n$, but when there are enough gossip phases (in particular, $g_p=5$), cooperation survives already for $n\geq3$.

\begin{figure}[h]
\centering
\includegraphics[width=6.5cm]{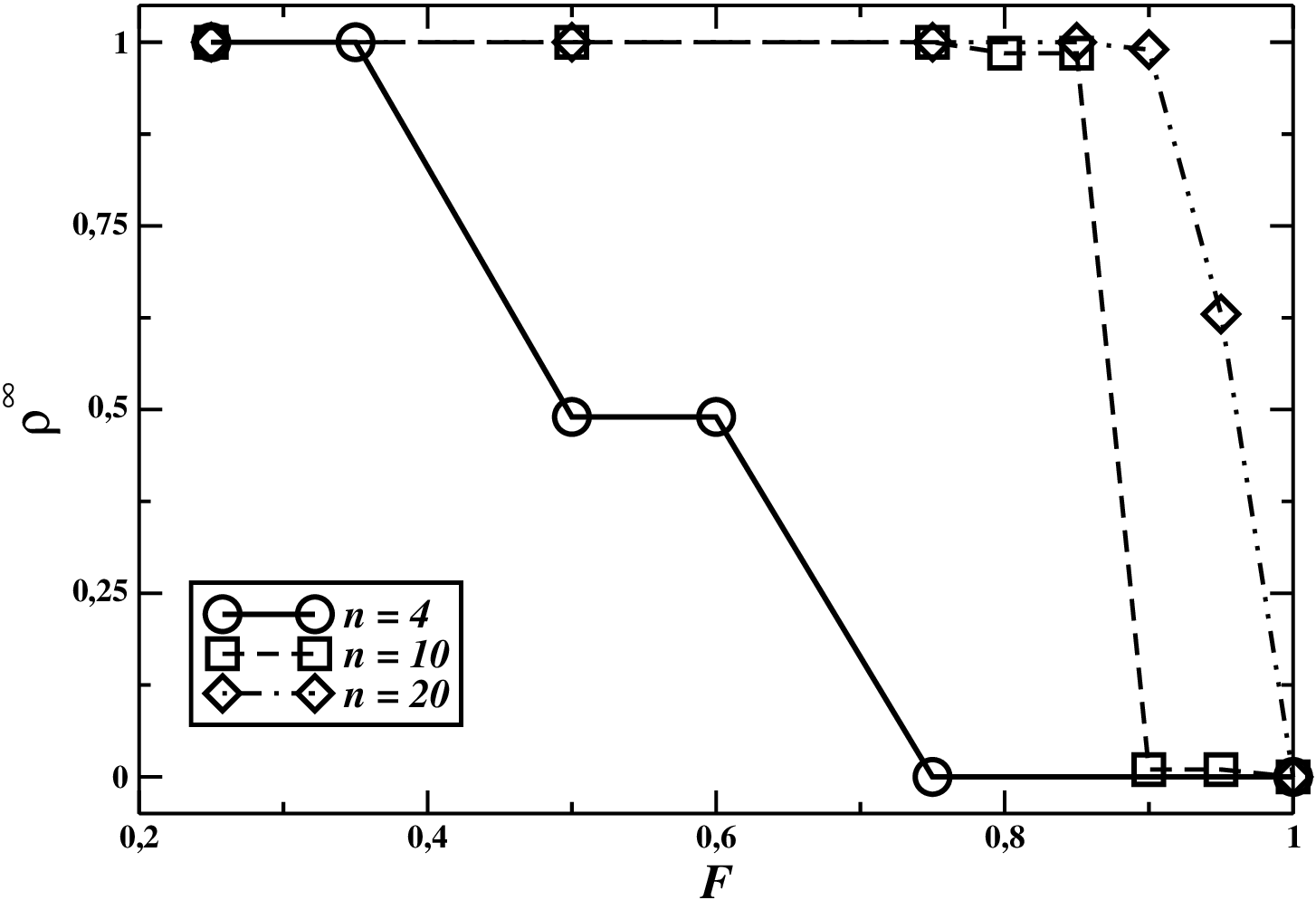}\qquad \qquad
\caption{
Final cooperator density as a function of the fraction $F$ of fixed members of the PGG groups for $g_p=5$, $n=4,\ 10$ and $20$, $M=100$ for $n=4$, $M=40$ for $n=10$, and $M=20$ for $n=20$, with gossip only among agents having good reputation of each other. Perfect information.}
\label{fin_coopF0.5}
\end{figure}

Finally, in the right graph of Figure~\ref{fin_coopF0.5} we show results for systems where only agents with non-negative reciprocal reputation exchange information. The behaviour looks identical to the one in the previous Figure, but the final cooperation decreases more rapidly with $F$ , and goes to zero with small or null partner selection for large group size. This means that also information spread by defectors can be useful for the emergence of cooperative behaviours at a global level, and keeping them away from gossip limits the emergence of pro-social strategies.

\ 

\begin{figure}[h]
  \centering
\includegraphics[width=6.5cm]{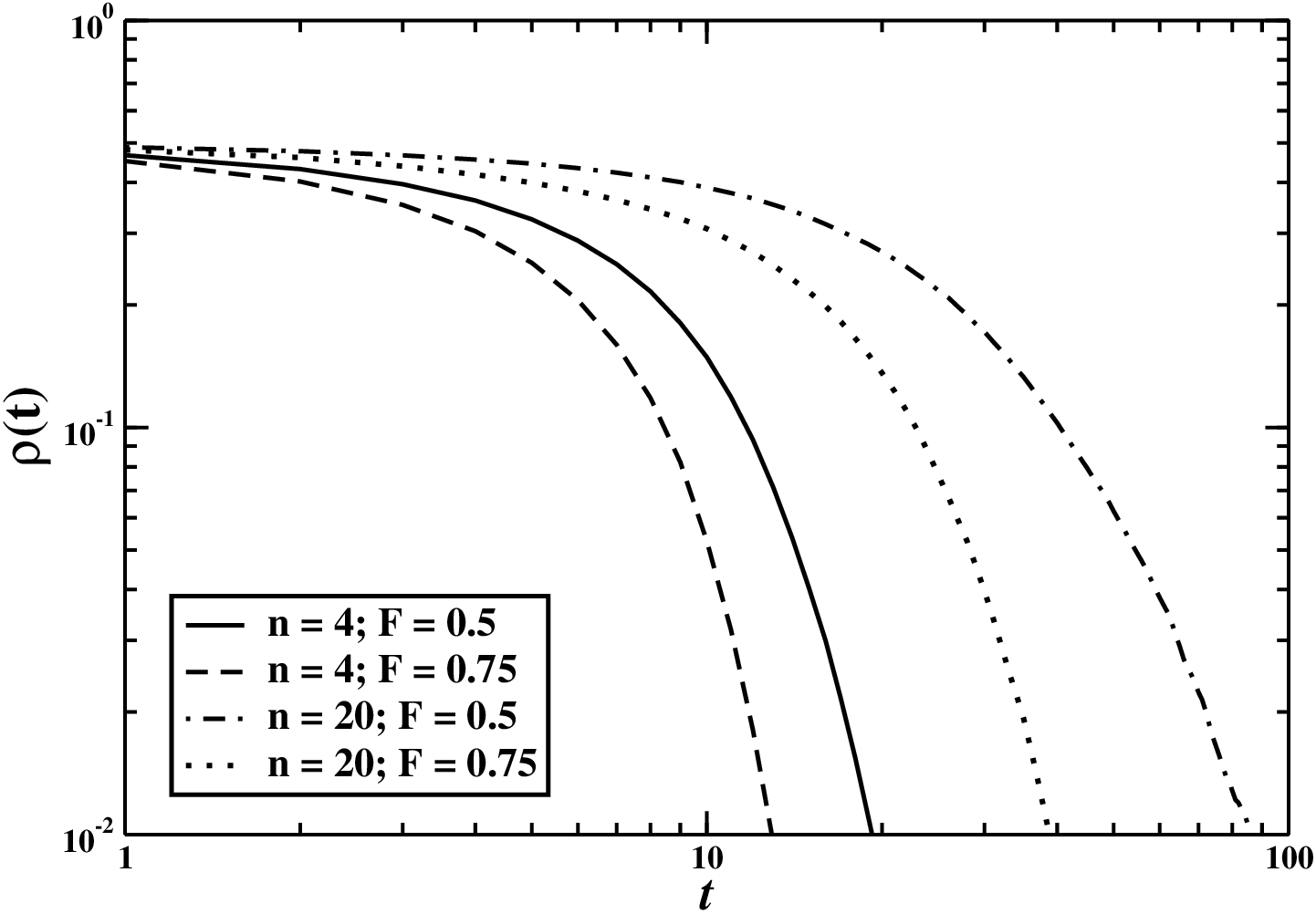}
\includegraphics[width=6.5cm]{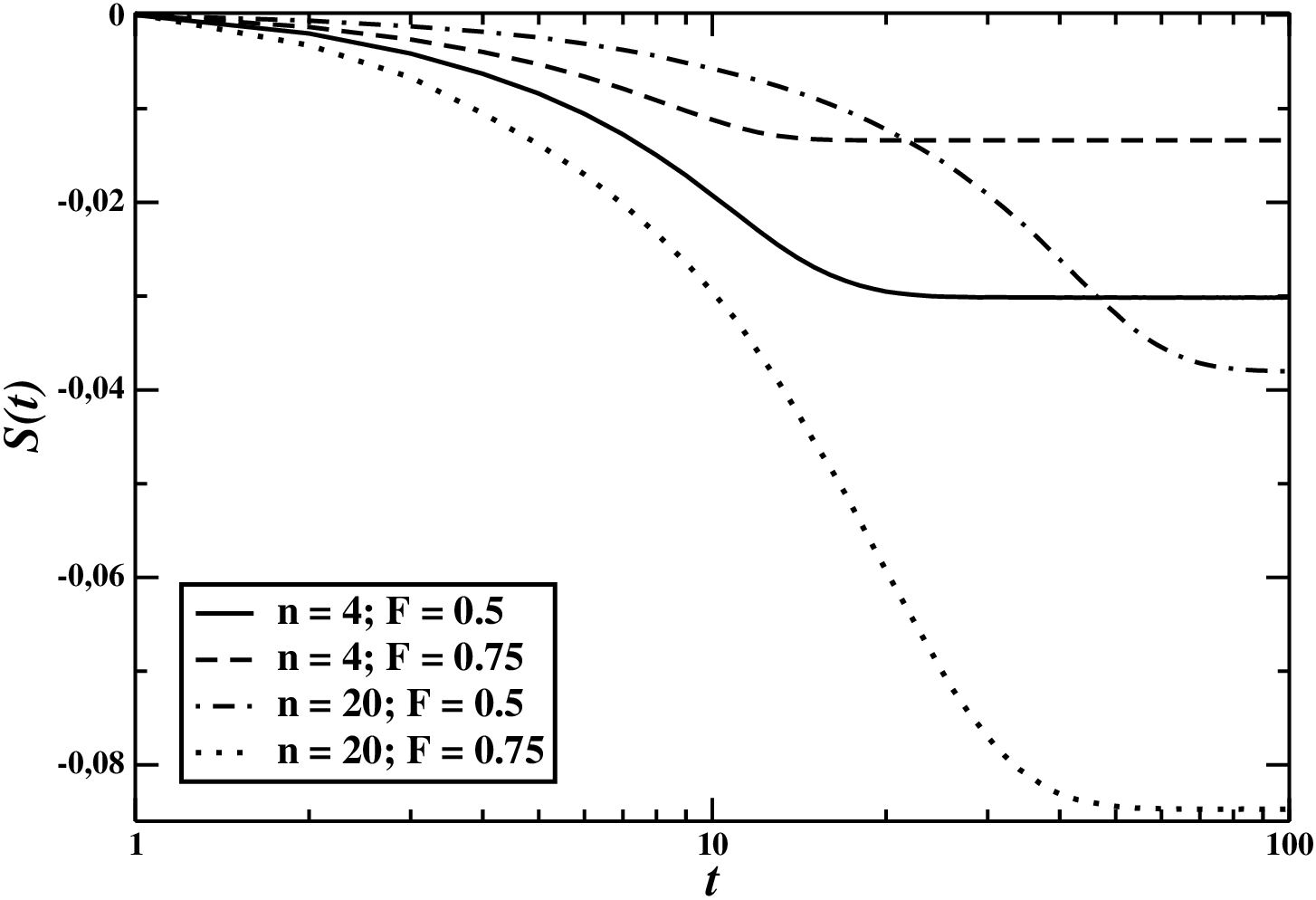}
\caption{
Time behaviour of cooperators density (left) and average reputation
(right) for $M=100$, $n=4,\ 20$, and $F=0.5,\ 0.75$ with no gossip effect.
Perfect information.}
\label{nogossip_fig}
\end{figure}

If we include the possibility of partner selection ({\it i.e.}, by setting $F<1$) without gossip ({\it i.e.}, keeping $g_p=1$), each group will be formed by $nF$ fixed members which choose the missing group members $n(1-F)$ at each round of the game.
As we can observe in Fig.~\ref{nogossip_fig}, also in this case cooperation goes to zero
at the end of the dynamics, even though less swiftly than in the previous modality with $F=1$.

\begin{figure}[h]
  \centering
\includegraphics[width=6.5cm]{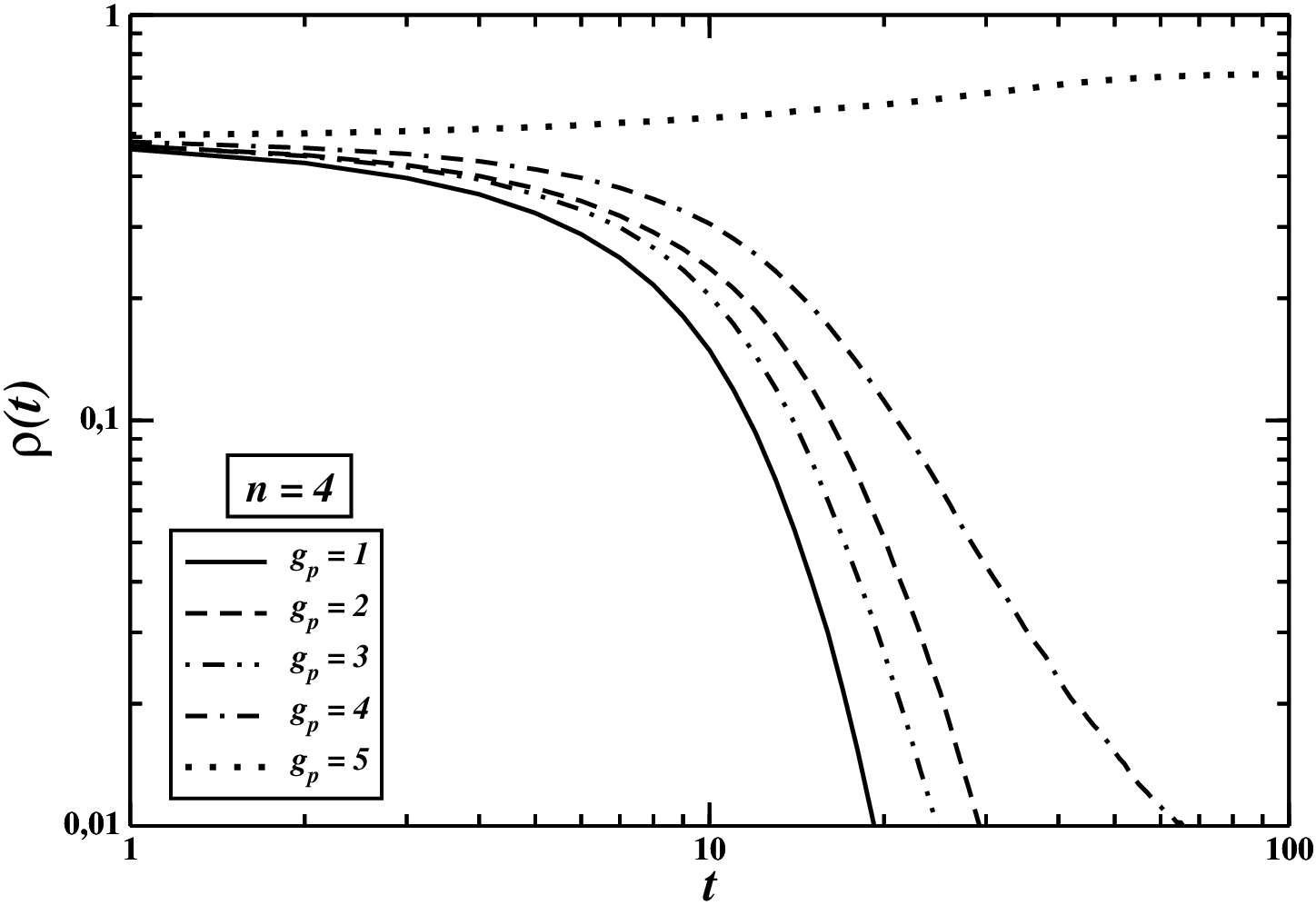}\qquad \qquad
\includegraphics[width=6.5cm]{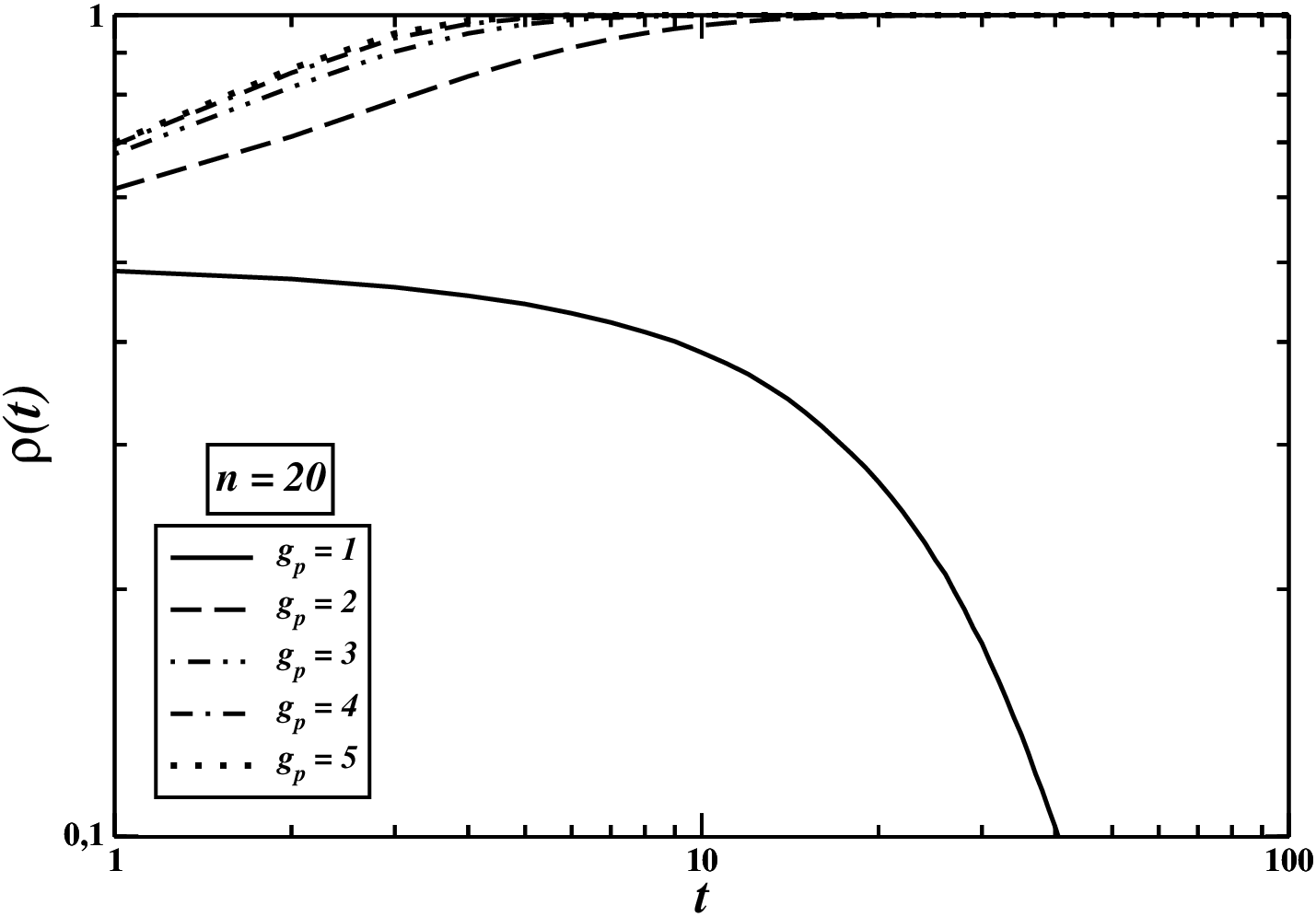}
\caption{
Time behaviour of the cooperator density for $n=4$ (left) and $n=20$ (right), with $L=200$, $F=0.5$ and different number $g_p$ of gossip phases. Perfect information.}
\label{coop_g04}
\end{figure}

The evolution of cooperation depends strongly on both $n$ and $g_p$, as shown in Figure~\ref{coop_g04}. Increasing both group size and the number of gossip stages makes cooperation thrive: the more individuals exchange information, the better they know each other so that they are able to avoid defectors when completing their own group in the PGG stages. Therefore, even though in a single PGG defectors might gain more than cooperators, if there is enough information flowing, defectors are quickly isolated, therefore they are accepted in fewer groups and gain lower payoffs.

\begin{figure}[h]
  \centering
\includegraphics[width=6.5cm]{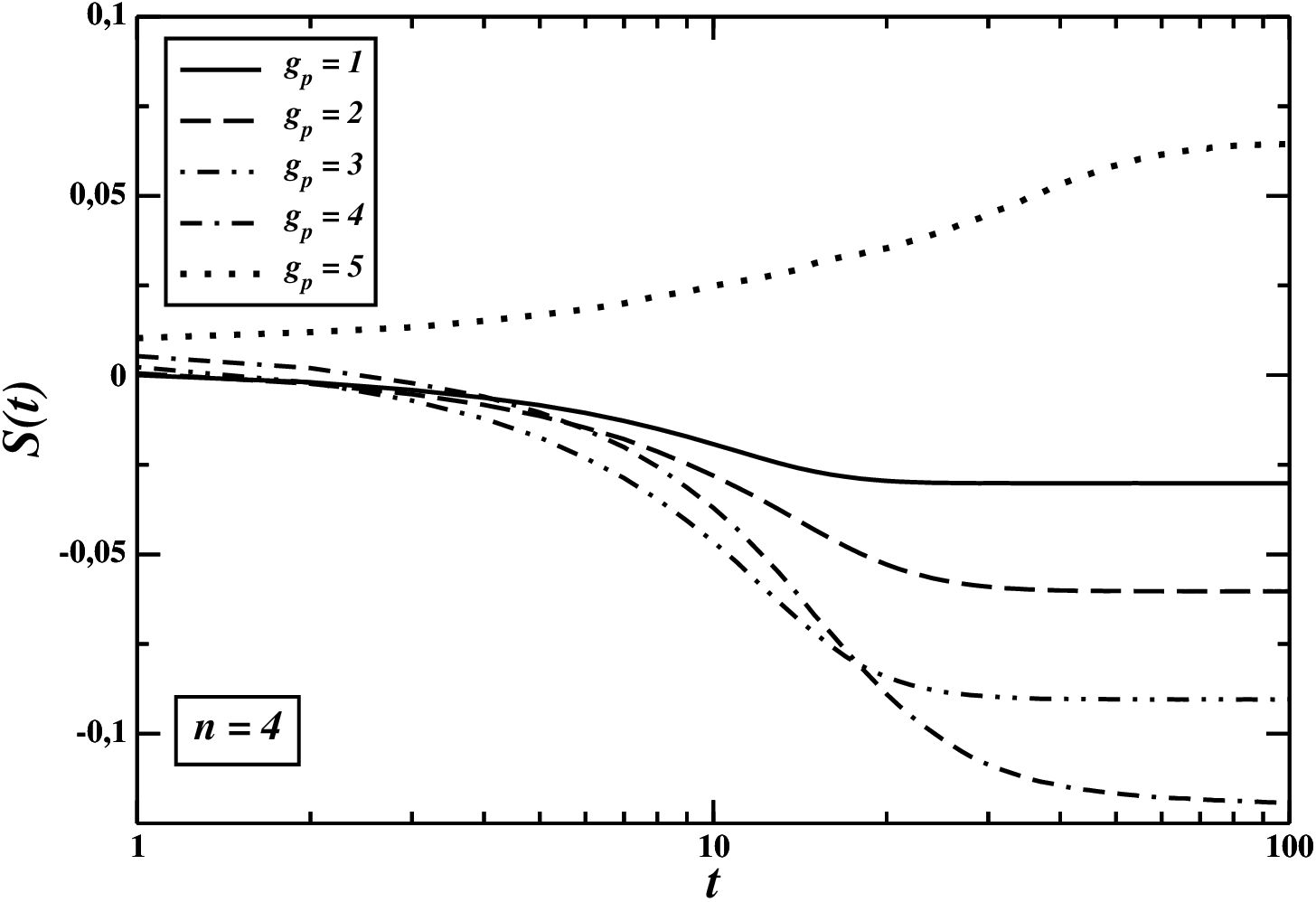}\qquad \qquad
\includegraphics[width=6.5cm]{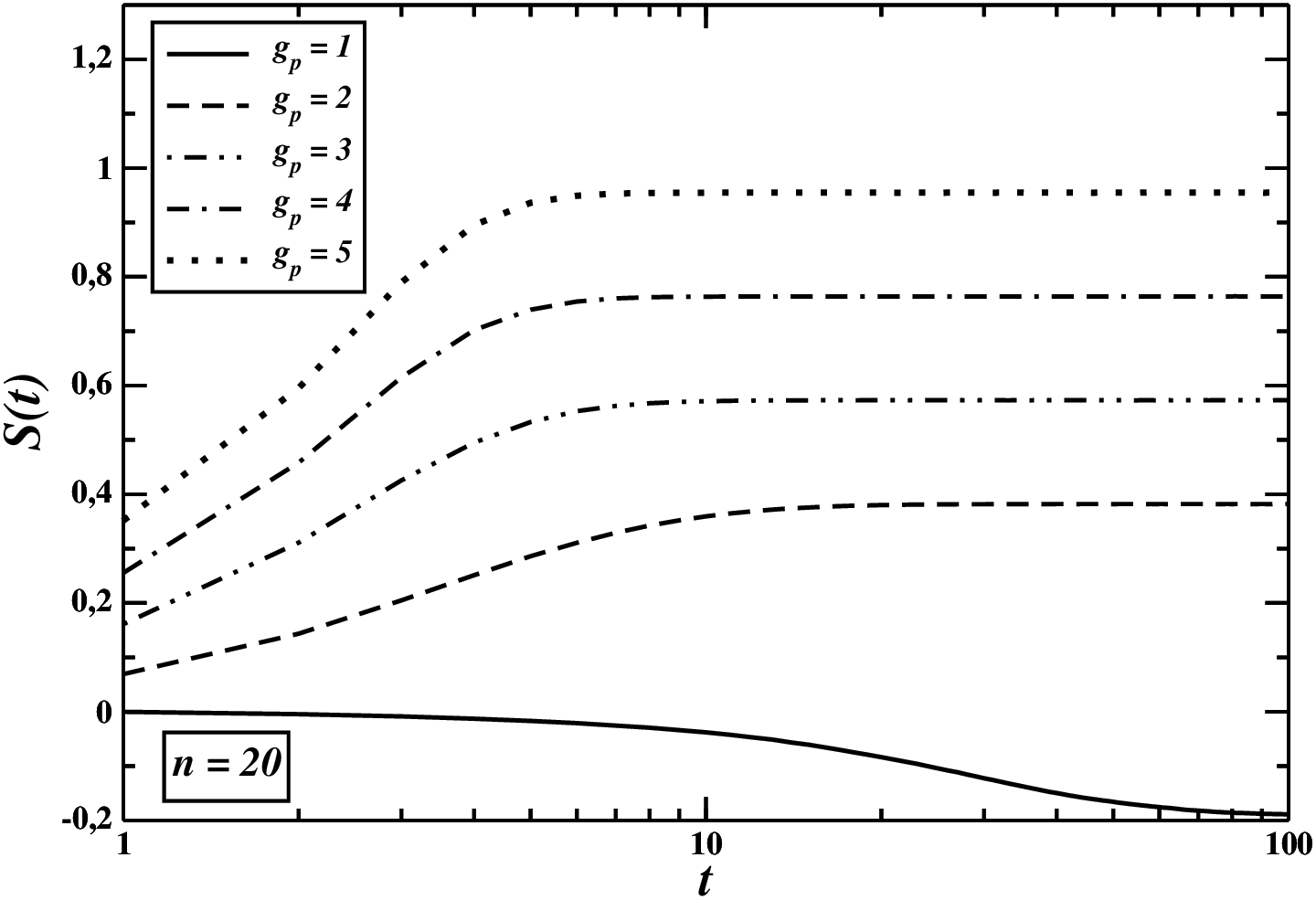}
\caption{
Time behaviour of the average reputation for $n=4$ (left) and $n=20$ (right), with $L=200$, $F=0.5$ and different number $g_p$ of gossip phases. Perfect information.}
\label{img_g04}
\end{figure}

The emergence of cooperation is mirrored also in average reputation. Average reputation converges to a negative value when cooperation gets extinct, and to a positive one when cooperators invade the system. Figures~\ref{nogossip_fig} (right graph) and~\ref{img_g04} show that the final average reputation increases when three conditions are met: if the number of gossip phases increases, if partner selection becomes harsher and, finally, if the group size also decreases. 

\begin{figure}[h]
  \centering
  \includegraphics[width=6.5cm]{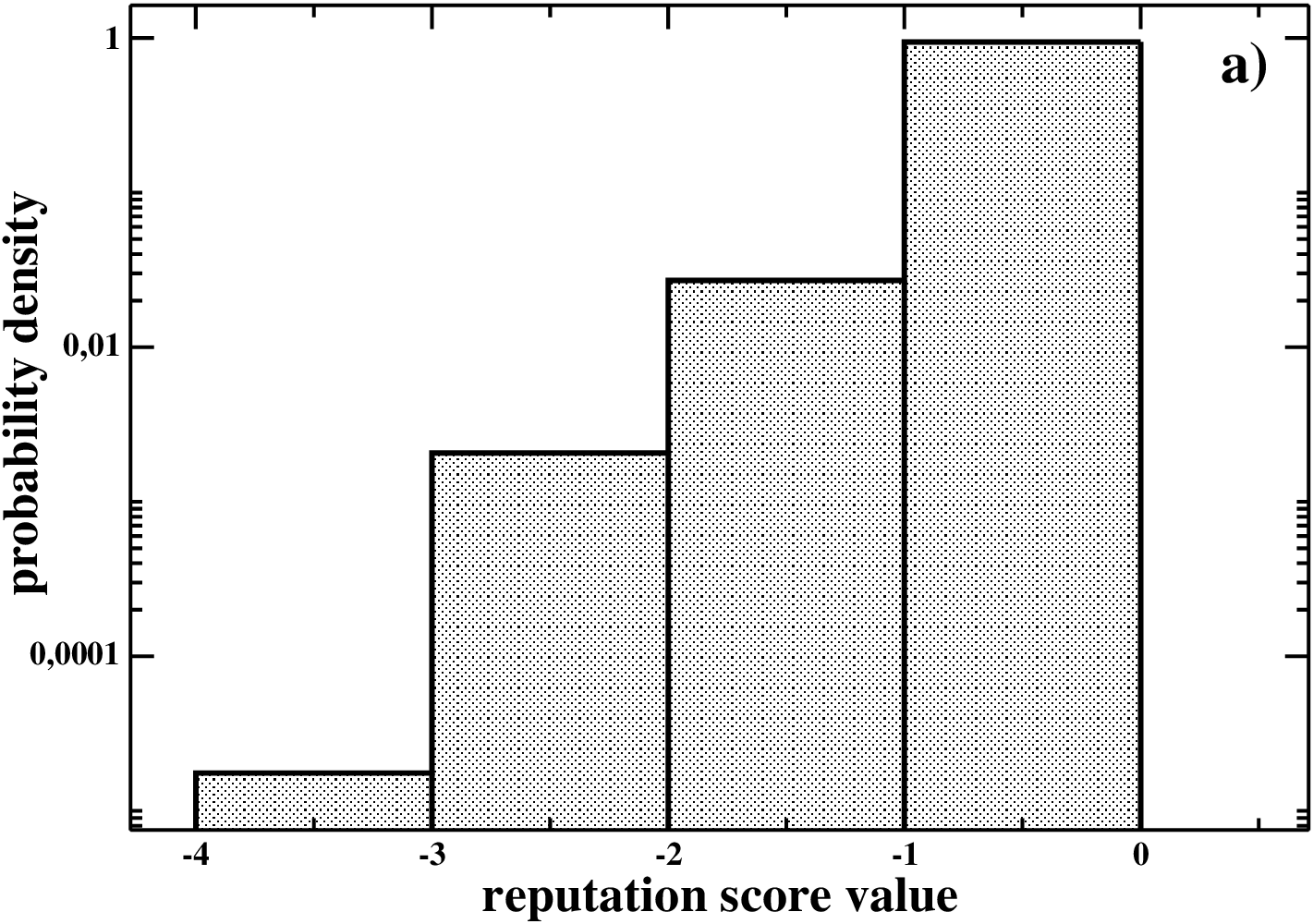}\qquad \qquad
  \includegraphics[width=6.5cm]{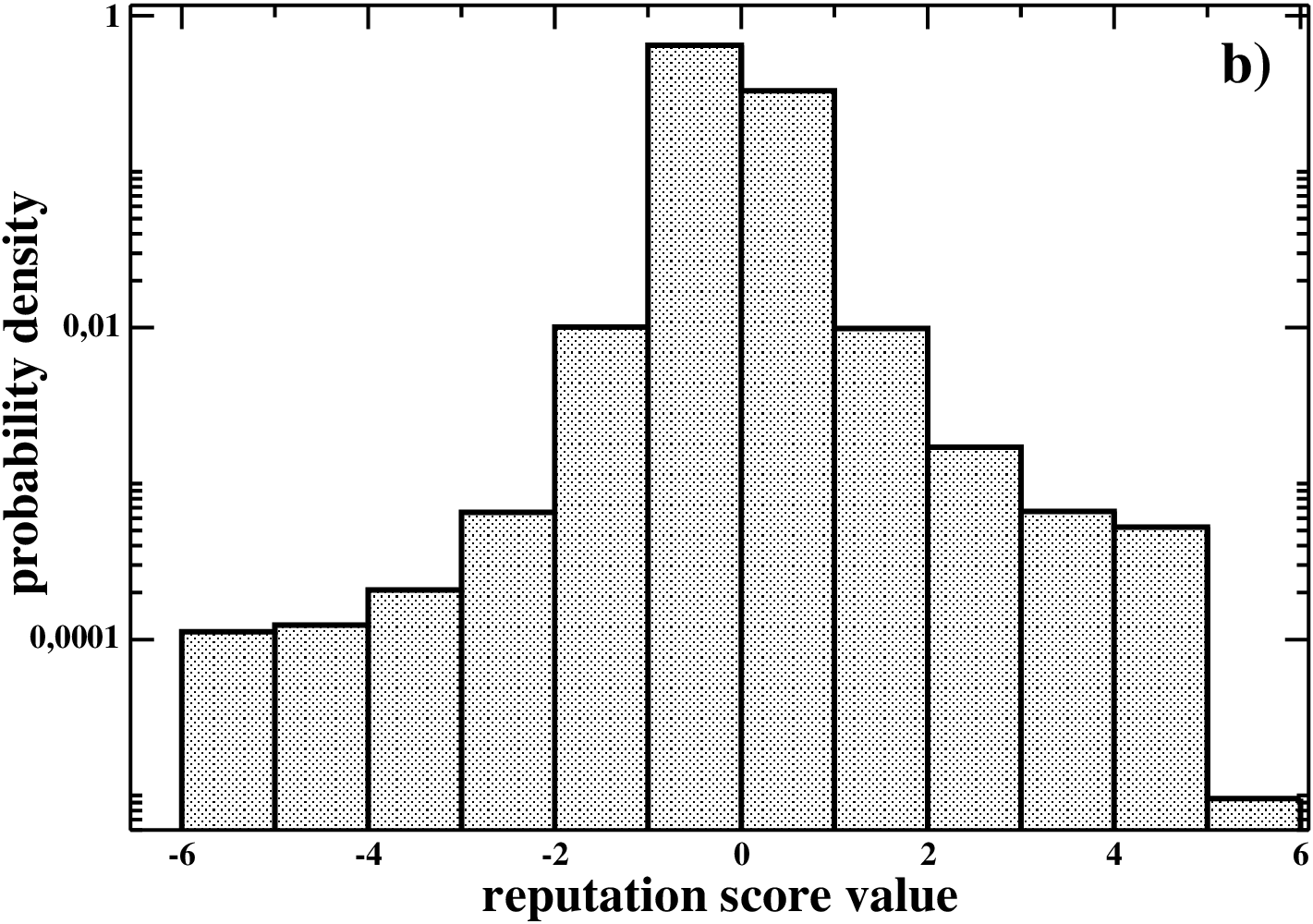}\qquad \qquad
  \includegraphics[width=6.5cm]{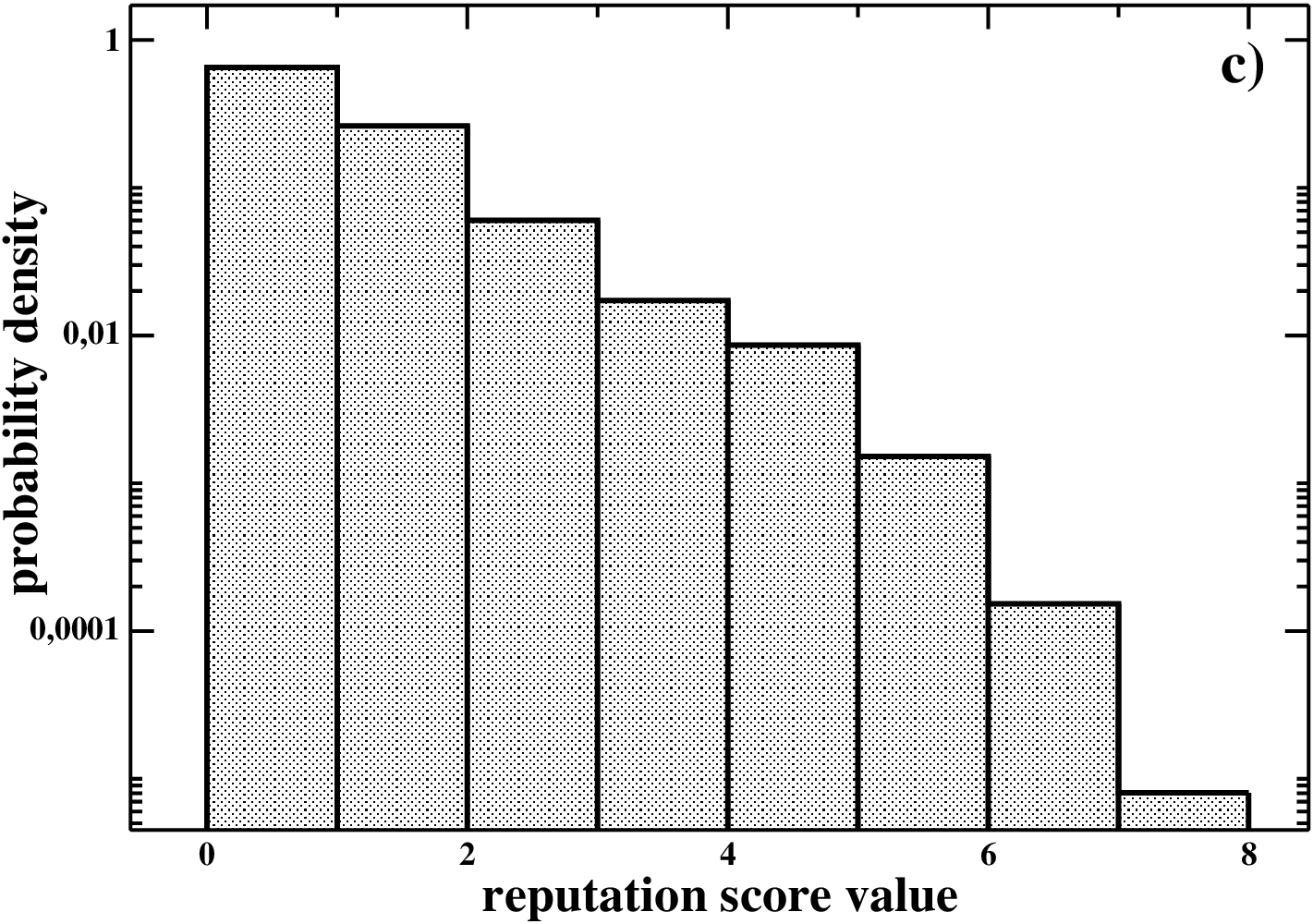}
\caption{
Final reputation score distributions for systems with $L=200$, $F=0.5$ and {\bf a)} $n=4$ and $g_p=2$,
{\bf b)} $n=4$ and $g_p=5$, {\bf c)} $n=20$ and $g_p=5$. Perfect information.}
\label{fin_sc_dist}
\end{figure}

Concerning the entire distributions of image score values, these are reported in Fig.~\ref{fin_sc_dist}. As we can observe, when cooperation disappears, every non-zero score is negative (of course, there are only defectors). At the same time, when cooperators invade completely the system the non-zero scores are always positive; when instead the final configuration is a mixed one, the final score distribution is closer to a symmetric one.

In summary, it is the combination of gossip quantity and partner selection harshness that makes cooperation survive. When there is not enough gossip to support partner selection ($g_p=1$), cooperation gets always extinct for any group size; on the other hand, as the number of gossip exchanges increases, the probability of invasion by cooperators increases as well.

\ 

The efficacy of gossip in supporting cooperation is limited by its reliability, but little is known about the effect of specific kinds of errors. In a situation in which the listener can misunderstand what the speaker says, there are two possible outcomes. An {\it exclusion error} occurs when, with probability $q_{\al}$, a gossip targets a player with positive reputation which is understood as negative by the listener. In a complementary way, an {\it inclusion error} refers to the probability $q_{\bt}$ that a player with negative reputation is mistakenly considered a cooperator.  

\ 

\begin{figure}[h]
  \centering
  \includegraphics[width=6.5cm]{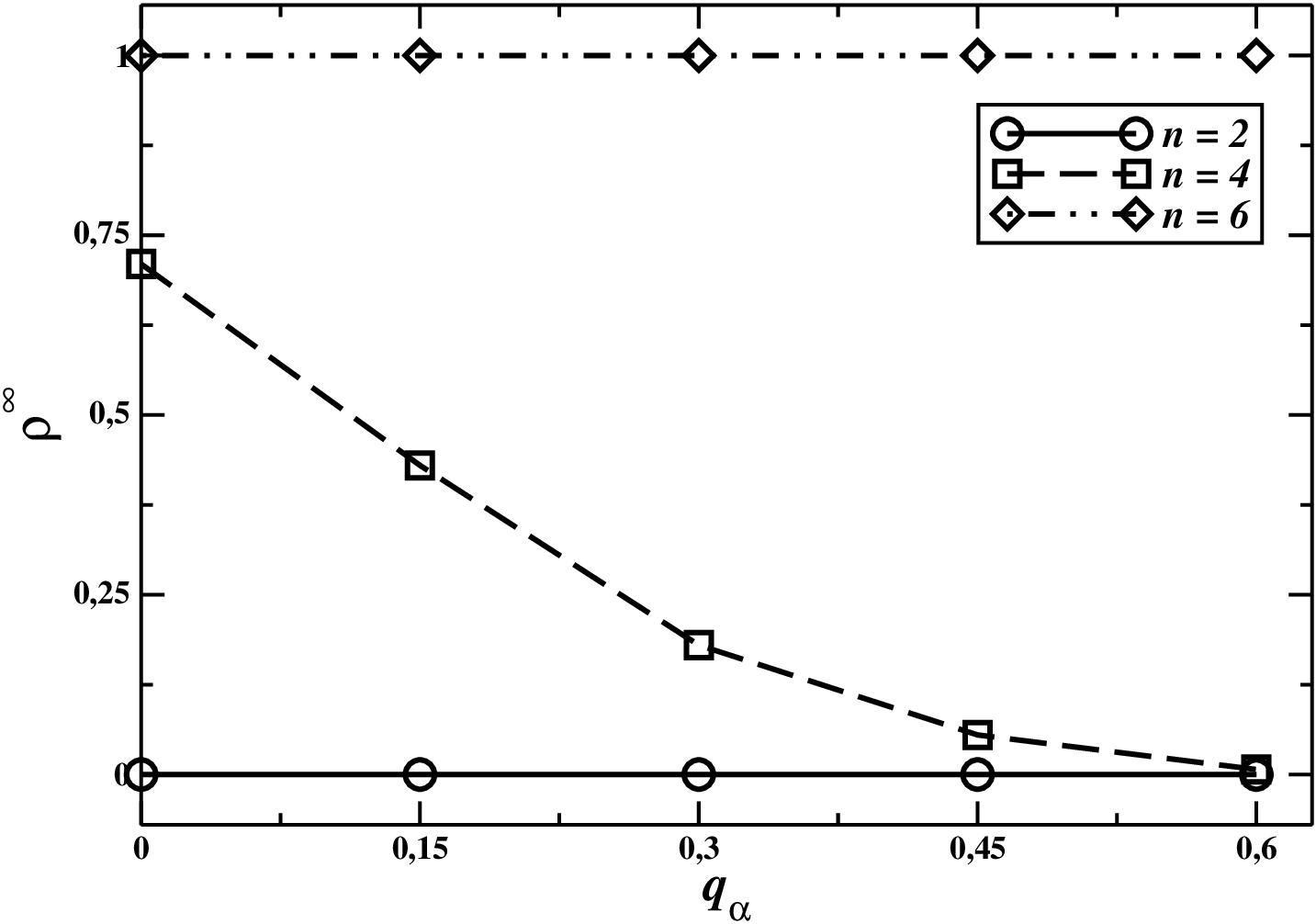}\qquad \qquad
  \includegraphics[width=6.5cm]{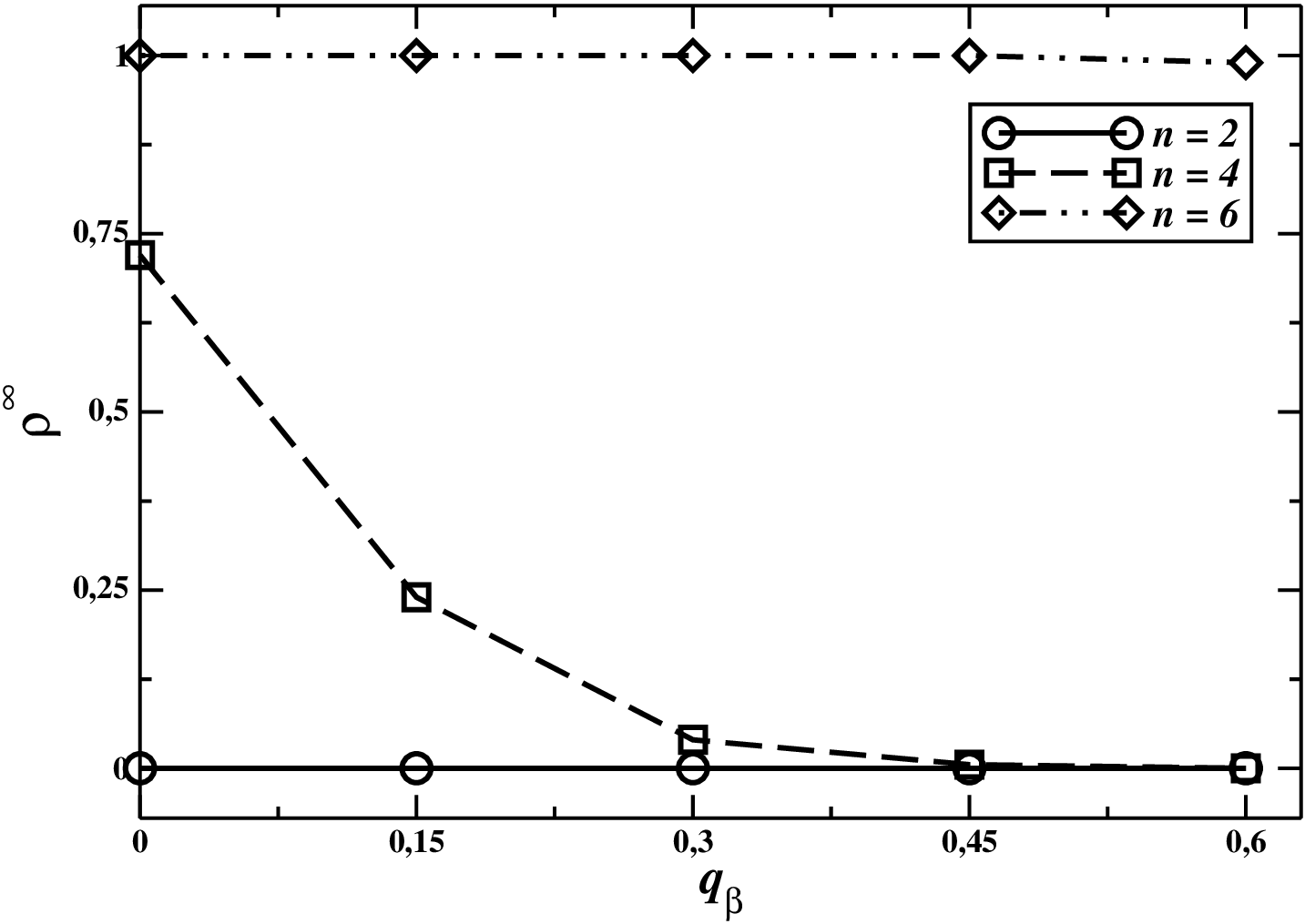}\qquad \qquad
  \includegraphics[width=6.5cm]{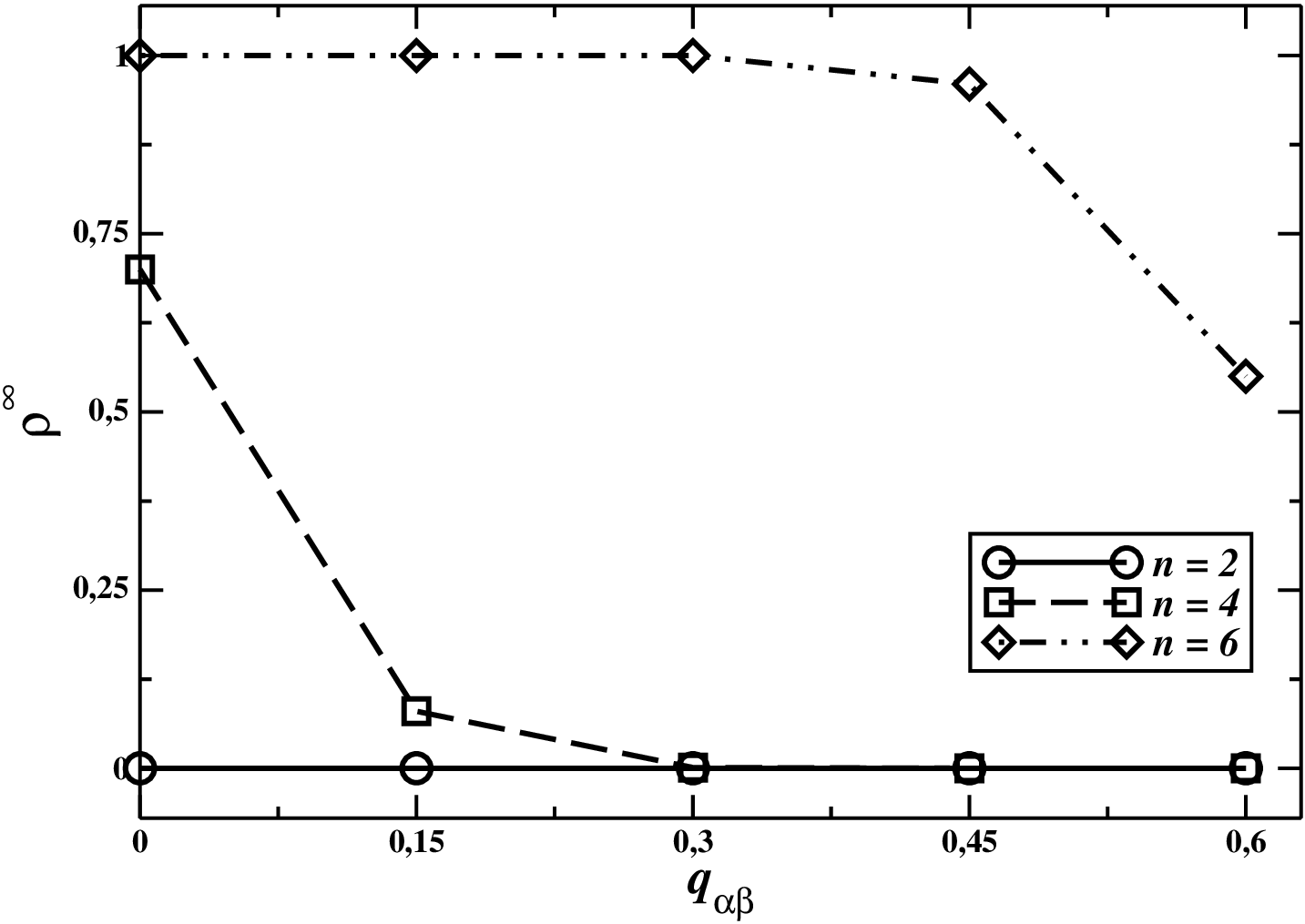}
\caption{
Final cooperator densities as a function of the transmission error probabilities $q_{\al}$ (top left), $q_{\bt}$ (top right), and $q_{\ab}$ (down) for $L=200$ ($L=201$ for $n=6$), $F=0.5$; $g_p=5$ and $n=2,\ 4,\ 6$. The case $n=2$, a PDG where the second player is accepted by the first one, is shown just as baseline. }
\label{fin_coopF0.5_q}
\end{figure}

With exclusion errors cooperators can still outcompete defectors for $n=20$, and they do not affect survival of contributors up to almost $q_\alpha=0.5$ for $n=4$ (Figure~\ref{fin_coopF0.5_q}, upper-left graph). This can be explained by the fact that in partner selection excluding defectors is essential, therefore, even though also some cooperators are excluded with $q_\alpha>0$, defectors are rejected as usual. In this
sense we could say that the Latin motto {\it "in dubio pro reo"} should be reversed ({\it "in
dubio in reum"}) in order to maintain high levels of cooperation in the population. Although not shown in the figure, we also verified that the time needed to reach the final state is of the same order of magnitude as in the case of perfect information transmission.

\ 

Inclusion errors take place when negative reputations are understood by the listener as positive, {\it i.e.}, a defector gains a positive reputation with probability $q_{\bt}$. Indeed, as we verified in our simulation, inclusion errors are more detrimental to cooperation with respect to the exclusion ones, since they allow defectors to be accepted by different groups despite their anti-social behaviour, thus lowering the global level of cooperation in the system. This is clear in the upper-right graph of Fig.~\ref{fin_coopF0.5_q}.

\ 

In this work we focused mainly in the case $q_{\al}q_{\bt}=0$ because our aim is to analyse separately the role and effect of each kind of errors. Anyway, in the real world errors (deliberate or not)
can happen in both ways. Therefore, we tested what happens when every information, no matter if positive or negative, can be wrongly transmitted with probability $q_{\al\bt}$. As reported in the lower graphics of Fig.~\ref{fin_coopF0.5_q}, we see that in this case the average final cooperation level is even lower than in the configuration with inclusion errors. This means that, if an exclusion error is not hindering the global cooperation because the key point is preventing defectors to be accepted in other groups, bidirectional errors have the worst global effects because they decrease the reliability of information overall. 

\ 

\section{Discussion}

Gossip is a key ingredient for the functioning of human societies, but its importance in evolutionary models of cooperation has been largely underrated. The present research contributes to our understanding of the role of gossip as an inexpensive but effective way of supporting cooperation in groups of varying size. 
Privately exchanged information helps cooperators finding and rejecting defectors, and this positive effect is reinforced when ostracism from the group is possible. Along with the structure of the interaction, the quantity of gossip plays a crucial role. The more information agents can exchange, the more accurate they become in selecting reliable partners, even when groups are unstable. 
Our results also show that the role of gossip abundance is modulated by group size: in small groups ($n=4$) at least five gossip stages are required to make cooperators outcompete defectors, whereas in large groups ($n=20$) two gossip stages are sufficient to collect and make good use of reported information. This finding complements results obtained in lab experiments where the abundance of gossip is effective in optimizing human responses, and in directing cooperation towards cooperators in an indirect-reciprocity games~\cite{som08}. 

Image score is an effective solution for supporting cooperation, but unlike previous work on image score in groups~\cite{nax15},
we designed a model in which agents exchange private information, showing that the combination of gossip and partner selection makes cooperation invade the system also for large sizes of the groups, whilst in absence of these mechanisms the final level of cooperation decreases with the group size~\cite{suz05,vil14}.

Gossip-based indirect reciprocity is expected to be not evolutionary stable if gossip is not completely reliable and accurate~\cite{now05}, but the size of this effect is debatable. This study  contributes to assess the importance of errors in transmission, showing that their effects are smaller than previously hypothesized, and that their disruptive power on reputation depends also on the kind of error. When unreliable information is introduced in the system, failing to exclude a defector is more detrimental to cooperation than erroneously ostracising a cooperator. This effect is amplified by the interaction structure, because in the bipartite graph defectors are accepted by different groups despite their anti-social behaviour. Even with excluding errors, cooperation emerges slowly, but in groups of 20 individuals it nonetheless goes to 1.

Unlike previous work~\cite{uch13,bra06,nak11}, we are not interested in errors in evaluations, which might consist in the application of the wrong social norm, but in transmission errors and their effects on groups' decision making. Pairing unreliable gossip with ostracism can lead either to ostracism against innocent cooperators or to acceptance of unrecognised defectors, and this work shows the different outcomes of these errors. 

A possible limitation of our study is that we do not consider malicious gossip strategically used to reduce someone's reputation. There is a good evidence that gossip includes both positive and negative talk and most interpersonal gossip is neutral or positive, as reported in different studies on every day conversations~\cite{lev85}. Duncan, Marriott, and Dunbar~\cite{dun97} report that less than 5 percent of the conversations they analysed had malicious and manipulative gossip as a topic. A second reason why we did not insert any manipulative gossip is because we wanted to proceed in a stepwise fashion, first testing whether privately exchanged information about an absent third party~\cite{eml94,fos04} could be effective in an evolutionary model of cooperation. However, strategic gossip is an interesting topic and we plan to address this limitation in our future work.
Another interesting development of our work could be in adding the costs of gossiping, that in real life are expressed in terms of potential punishment by those who are gossiped about~\cite{gia12b,szo15}. 

Another crucial element for supporting cooperation is the interaction between group-size and a bipartite graph, whose combination is very effective in isolating defectors. According to Nowak and May~\cite{now92}, in computational models of the evolution of cooperation, the structure of interactions among individuals could be important in enhancing cooperation even though at the individual level this strategy results detrimental. A great deal of studies on evolutionary game theory on graphs were spawned by this first, key insight~\cite{sza07,tom07,roc09,vil11}, making it clear that many factors can favour, or hinder, global cooperative behaviours. When agents playing a PGG are placed on a bipartite graph with image score opportunities~\cite{suz05}, cooperation can emerge and be maintained, also for different group sizes ~\cite{vil14}. When cooperation is framed as a Public Goods Game (PGG)~\cite{har68,gar90}, cooperation can hardly be sustained, unless costly punishment is provided~\cite{car07,feh99}. Although effective in many contexts, punishment increases the amount of cooperation but not the average pay-off for the group~\cite{dre08}, and in repeated games cooperators who do not bear the costs of punishing defectors are better off than cooperators who punish~\cite{oht09}. Placing agents on a bipartite graph allows us to test the effects of ostracism as a low-cost and effective form of punishment, but also to understand whether a combination of gossip and ostracism might support the evolution of cooperation. 

Early small-scale experiments suggested that network topology may actually enhance cooperative strategies in controlled laboratory situations~\cite{cas07,sur11,tra10}. Though, larger-scale and more in-depth studies and reviews~\cite{hel10,gru10,gra12,gru14} showed that there is no significant influence of the interaction network on the emergence and evolution of cooperation in behavioural experiments using a Prisoner's Dilemma Game (PDG). Analogously, it was analytically demonstrated that bipartite networks with PGG are very effective in fostering pro-social behaviours~\cite{gar11}.

By linking the results on gossip, in terms of both quantity and quality, with the data on ostracism on a bipartite graph, we provide support for the role of gossip-based indirect reciprocity on the evolution of cooperative behaviours in groups. This finding has implications for the current debate on the evolution of cooperation, showing that more realistic mechanisms, like gossip and ostracism, can be as effective as more ideal-typical image score. 

\ 

\section{Methods}
\label{model}

We consider a population of $L$ individuals placed on a bipartite graph~\cite{ram04,die97,gui05,pel06,gar11}, that is, a network containing two kinds of nodes denoting agents and groups, respectively. This implies that links can be established only between nodes of different types whereas no direct connection among individual agents is
possible. An overview of the main parameters is presented in Table~\ref{table1}.
\ 

{\it Network building - } The network building proceeds as follows: given a value $F\in(0,1)$, we set $nF$ initial members for each group so that each individual belongs exclusively to one group. As an illustration, consider the case with $L=150$, $n=20$ and $F=0.75$ (then $M=10$). This would mean that we have 15 agents in the first group, other
15 in the second one and so on until the last group is formed. Then, each group needs to be completed with a subset $(1-F)n=5$ of individuals selected from a pool of available candidates. Potential partners are randomly selected from the remaining population, but they can become members of that group if and only if group members have, on average, a non-negative reputation $\sigma$ of potential candidates. If the number of available partners (that is, candidates with non-negative average reputation) is not enough to complete the group, it will be filled by random selection of new members.
Reputations are expressed by means of integers: in general, $\sigma(x,y)$ is the reputation that the individual $y$ has in $x$'s eyes. Whenever an agent $i$ interacts with an agent $j$ during a PGG stage, their respective reputations are updated. If $j$ has cooperated, $\sigma(i,j)$ is increased by 1, otherwise is decreased by 1. This process is bilateral, thus 
the same happens to the other agent's record $\sigma(j,i)$ of $j$ about $i$. There is no maximum nor minimum limit to the possible values that $\sigma(i,j)$ can assume. This direct experience is also complemented by information exchanged during gossip stages (for details, see the description of the  Gossip stage below). If there is nobody in the whole population with a non-negative average reputation, new members are accepted regardless of their standing.

$L$ individuals are distributed into $M$ groups, each group composed of $n$ members playing a sequence of PGG and gossip phases. Each and every individual $i$ is characterized by an innate 
strategy $s_i$, which can be either cooperation (C), or defection (D). 
Moreover, each agent has private information about other individuals' reputation, on the basis of direct interaction and gossip.

\

{\it Initial conditions - } At the onset of the simulation, strategies are randomly assigned to players, therefore we have on average
$50\%$ of cooperators and $50\%$ of defectors. Reputations are set equal to zero: $\sigma^{in}(i,j)=0\ \forall i,j$.

\

{\it Dynamics - } The dynamics consist of $g_p$ phases, each phase characterized by the combination of a game round followed by a gossip round. In PGG rounds, cooperators contribute a quantity $c>0$, whereas defectors contribute nothing. The total contribution in each group is multiplied by a factor $nB$ and equally shared among all group members, regardless of individual contributions. Each agent plays as many PGG rounds as the number of groups it belongs to, and the total fitness of a player is the sum of the pay-off gained in each of its groups, with $B=0.85$ and $c=1$ as in the work by Suzuki and Akiyama~\cite{suz05}.
  
  \ 

 {\it Gossip stage - } Reputations are privately held beliefs that agents update on the basis of their direct observation and, when gossip is available, of the information they receive from another agent.
 Therefore, in the gossip stage, agents interact in pairs where the first one, $i$, acts as the ``speaker'' and the second, $j$, as the ``listener''. The target of gossip is a third player $l$ whose reputation $\sigma(i,l)$ is communicated by $i$ to $j$. The reputation of $l$ is updated by averaging $j$'s original knowledge with the newest:
  
\be
\sigma(j,l)\rightarrow\sigma_{new}(j,l)=[\sigma(j,l)+\sigma(i,l)]/2 \ . 
\label{gossip1}
\ee
  
{\it Imperfect information - }  Since information transmission is prone to errors of several origins going from noise to opportunistic deception, we enriched the model by including two different kinds of errors. The first ones are called {\it exclusion} errors, which make cooperative individuals have a negative reputations, and {\it inclusion} errors, which make defectors have positive reputations.
Exclusion errors are implemented as follows: if $\sigma(i,l)>0$, that is, if $l$ has a good reputation according to $i$, there is a probability $q_{\al}$ that reputation transmission is wrong and the receiver, $j$ understands the algebraic opposite of what is told, so that

\be 
\sigma(j,l)\rightarrow\sigma_{new}(j,l)=[\sigma(j,l)-\sigma(i,l)]/2 \ . 
\label{gossip2}
\ee
   
On the contrary, we have an inclusion error when $l$ has a negative reputation to $i$, and there is a probability $q_{\bt}$ that $j$ understands $l$ is a cooperator, so that

\be
\sigma(j,l)\rightarrow\sigma_{new}(j,l)=[\sigma(j,l)+|\sigma(i,l)|]/2 \ . 
\label{gossip3}
\ee
   
  In this work we assume that $q_{\al}\cdot q_{\bt}=0$, that is, 
  both can be equal to zero but at most one of them can be positive. 
  This procedure is repeated $2\nu L$ times, so that each player
  happens to be on average $\nu$ times a speaker and $\nu$ times a listener. Here, we always assume $\nu=50$. Moreover, in order to investigate the full range of effects of errors, we also tested bidirectional errors, where with probability $q_{\al\bt}$ the same process described in Eq.~\ref{gossip2} is applied.

  New group members are selected before each of the $g_p$ game rounds.
  It is important to stress that, if $g_p=1$, then gossip has no effect on the dynamics of the population.

\

{\it Reproduction - } After $g_p$ game-gossip phases, the reproduction stage takes place. Reproduction is modelled as binary tournament selection~\cite{suz05}. Two individuals are randomly selected from the overall population (parents) and a new individual (offspring) is created. The offspring inherits the strategy of the parent with the highest fitness with probability $P=0.9$ , otherwise the strategy of the less performing parent is inherited. The parents are put again in the original population, and offspring is stored in another pool.
When this selection process has happened $L$ times, the old population is deleted and replaced with the offspring. The offspring inherits only the parents' strategy, while the reputations $\sigma(i,j)$ are again set to zero: in this respect, we followed again the work of Nowak and Sigmund~\cite{now98a}, and Suzuki and Akiyama~\cite{suz05}. On the other hand, models which consider more complex evolution mechanisms have been considered for simpler interactions, as for example populations playing the Prisoner's Dilemma Game in co-evolving networks~\cite{szo09b,szo09c}.

Each new generation repeats the $g_p$ game-gossip phases, after which another reproduction stage takes place.
The simulation goes on for $t$ generations, until a final steady configuration is reached by the system. Depending on the parameter values chosen, the number of generations needed to reach such final state can go from ten to about one hundred.

\ 

\section*{Acknowledgements}

DV acknowledges support from H2020 FETPROACT-GSS CIMPLEX Grant No. 641191.

\section*{Author contributions statement}

F.G. conceived the model,  D.V. conducted the simulations, F.G. and D.V. analysed the results and wrote the manuscript. Both authors reviewed the manuscript. 

\ 

\bibliographystyle{plain}
\bibliography{GossipArticle2016b_arXiv}

\begin{thebibliography}{10}

\bibitem{ble76}
Wolf Bleek.
\newblock Witchcraft, gossip and death: A social drama.
\newblock {\em Man}, pages 526--541, 1976.

\bibitem{bra06}
Hannelore Brandt and Karl Sigmund.
\newblock The good, the bad and the discriminator—errors in direct and
  indirect reciprocity.
\newblock {\em Journal of theoretical biology}, 239(2):183--194, 2006.

\bibitem{bur08}
Ronald~S Burt.
\newblock Gossip and reputation.
\newblock {\em Management et r{\'e}seaux sociaux: ressource pour l’action ou
  outil de gestion}, pages 27--42, 2008.

\bibitem{car07}
Jeffrey~P Carpenter.
\newblock Punishing free-riders: How group size affects mutual monitoring and
  the provision of public goods.
\newblock {\em Games and Economic Behavior}, 60(1):31--51, 2007.

\bibitem{cas07}
A.~Cassar.
\newblock Coordination and cooperation in local, random and small world
  networks: Experimental evidence.
\newblock {\em Games and Economic Behavior}, 58:209, 2007.

\bibitem{cin05}
Matthias Cinyabuguma, Talbot Page, and Louis Putterman.
\newblock Cooperation under the threat of expulsion in a public goods
  experiment.
\newblock {\em Journal of public Economics}, 89(8):1421--1435, 2005.

\bibitem{die97}
R.~Diestel.
\newblock {\em Graph Theory}.
\newblock Springer, 4 edition, 1997.

\bibitem{dre08}
Anna Dreber, David~G Rand, Drew Fudenberg, and Martin~A Nowak.
\newblock Winners don’t punish.
\newblock {\em Nature}, 452(7185):348--351, 2008.

\bibitem{dun04}
Robin~IM Dunbar.
\newblock Gossip in evolutionary perspective.
\newblock {\em Review of general psychology}, 8(2):100, 2004.

\bibitem{dun97}
Robin~IM Dunbar, Anna Marriott, and Neil~DC Duncan.
\newblock Human conversational behavior.
\newblock {\em Human Nature}, 8(3):231--246, 1997.

\bibitem{eml94}
Nicholas Emler.
\newblock {\em Good gossip}, chapter Gossip, reputation, and social adaptation.
\newblock Univ Pr of Kansas, Kansas, 1994.

\bibitem{feh99}
Ernst Fehr and Simon G{\"a}chter.
\newblock Cooperation and punishment in public goods experiments.
\newblock {\em Institute for Empirical Research in Economics working paper},
  (10), 1999.

\bibitem{fei14}
Matthew Feinberg, Robb Willer, and Michael Schultz.
\newblock Gossip and ostracism promote cooperation in groups.
\newblock {\em Psychological Science}, 25(3):656--664, 2014.

\bibitem{fos04}
Eric~K Foster.
\newblock Research on gossip: Taxonomy, methods, and future directions.
\newblock {\em Review of General Psychology}, 8(2):78, 2004.

\bibitem{gar90}
Roy Gardner, Elinor Ostrom, and James~M Walker.
\newblock The nature of common-pool resource problems.
\newblock {\em Rationality and Society}, 2(3):335--358, 1990.

\bibitem{gia12b}
Francesca Giardini.
\newblock Deterrence and transmission as mechanisms ensuring reliability of
  gossip.
\newblock {\em Cognitive processing}, 13(2):465--475, 2012.

\bibitem{gia12}
Francesca Giardini and Rosaria Conte.
\newblock Gossip for social control in natural and artificial societies.
\newblock {\em Simulation}, 88(1):18--32, 2012.

\bibitem{glu63}
Max Gluckman.
\newblock Papers in honor of melville j. herskovits: Gossip and scandal.
\newblock {\em Current anthropology}, 4(3):307--316, 1963.

\bibitem{glu68}
Max Gluckman.
\newblock Psychological, sociological and anthropological explanations of
  witchcraft and gossip: A clarification.
\newblock {\em Man}, 3(1):20--34, 1968.

\bibitem{gar11}
J.~G\'omez-Garde\~nes, M.~Romance, R.~Criado, D.~Vilone, and A.~S\'anchez.
\newblock Evolutionary games defined at the network mesoscale: The public goods
  game.
\newblock {\em Chaos}, 21:016113, 2011.

\bibitem{gra12}
C.~Gracia-L\'azaro, A.~Ferrer, G.~Ruiz, A.~Taranc\'on, J.~A. Cuesta, and
  A.~S\'anchez.
\newblock Heterogeneous networks do not promote cooperation when humans play a
  prisoner’s dilemma.
\newblock {\em Proceedings of National Academy of Sciences USA},
  109:12922:12926, 2012.

\bibitem{gru10}
J.~Gruji\'c, C.~Fosco, L.~Araujo, J.~A. Cuesta, and A.~S\'anchez.
\newblock Social experiments in the mesoscale: Humans playing a spatial
  prisoner's dilemma.
\newblock {\em PLoS ONE}, 5:e13749, 2010.

\bibitem{gru14}
J.~Gruji\'c, C.~Gracia-L\'azaro, M.~Milinski, D.~Semmann, A.~Traulsen, J.~A.
  Cuesta, Y.~Moreno, and A.~S\'anchez.
\newblock A comparative analysis of spatial prisoner’s dilemma experiments:
  Conditional cooperation and payoff irrelevance.
\newblock {\em Scientific Reports}, 4:4615, 2014.

\bibitem{gui05}
R.~Guimer\`a, B.~Uzzi, J.~Spiro, and L.~A.~N. Amaral.
\newblock Team assembly mechanisms determine collaboration network structure
  and team performance.
\newblock {\em Science}, 308:697, 2005.

\bibitem{har68}
Garrett Hardin.
\newblock The tragedy of the commons.
\newblock {\em Science}, 162(3859):1243--1248, 1968.

\bibitem{hel10}
D.~Helbing and W.~Yu.
\newblock The future of social experimenting.
\newblock {\em Proceedings of National Academy of Sciences USA}, 107:5265,
  2010.

\bibitem{lev85}
Jack Levin and Arnold Arluke.
\newblock An exploratory analysis of sex differences in gossip.
\newblock {\em Sex Roles}, 12(3-4):281--286, 1985.

\bibitem{mai05}
Frank~P Maier-Rigaud, Peter Martinsson, and Gianandrea Staffiero.
\newblock Ostracism and the provision of a public good-experimental evidence.
\newblock {\em MPI Collective Goods Preprint}, (2005/24), 2005.

\bibitem{vaq13}
L.~A. Martinez-Vaquero and J.~A. Cuesta.
\newblock Evolutionary stability and resistance to cheating in an indirect
  reciprocity model based on reputation.
\newblock {\em Physical Review E}, 84(5):052810, 2013.

\bibitem{nak11}
Mitsuhiro Nakamura and Naoki Masuda.
\newblock Indirect reciprocity under incomplete observation.
\newblock {\em PLOS Comput Biol}, 7(7):e1002113, 2011.

\bibitem{nax15}
Heinrich~H Nax, Matja{\v{z}} Perc, Attila Szolnoki, and Dirk Helbing.
\newblock Stability of cooperation under image scoring in group interactions.
\newblock {\em Scientific reports}, 5, 2015.

\bibitem{new93}
M.~E.~J. Newman.
\newblock The structure and function of complex networks.
\newblock {\em SIAM Review}, 45:167, 2003.

\bibitem{new04}
Mark~EJ Newman.
\newblock Analysis of weighted networks.
\newblock {\em Physical Review E}, 70(5):056131, 2004.

\bibitem{now98b}
M.~Nowak and K.~Sigmund.
\newblock The dynamics of indirect reciprocity.
\newblock {\em Journal of Theoretical Biology}, 194:561--574, 1998.

\bibitem{now98a}
M.~Nowak and K.~Sigmund.
\newblock Evolution of indirect reciprocity by image scoring.
\newblock {\em Nature}, 393:573--577, 1998.

\bibitem{now92}
M.~A. Nowak and R.~M. May.
\newblock Evolutionary games and spatial chaos.
\newblock {\em Nature}, 359:826, 1992.

\bibitem{now05}
Martin~A Nowak and Karl Sigmund.
\newblock Evolution of indirect reciprocity.
\newblock {\em Nature}, 437(7063):1291--1298, 2005.

\bibitem{oht06}
Hisashi Ohtsuki and Yoh Iwasa.
\newblock The leading eight: social norms that can maintain cooperation by
  indirect reciprocity.
\newblock {\em Journal of Theoretical Biology}, 239(4):435--444, 2006.

\bibitem{oht09}
Hisashi Ohtsuki, Yoh Iwasa, and Martin~A Nowak.
\newblock Indirect reciprocity provides only a narrow margin of efficiency for
  costly punishment.
\newblock {\em Nature}, 457(7225):79--82, 2009.

\bibitem{ouw05}
Jaap~W Ouwerkerk, Norbert~L Kerr, Marcello Gallucci, and Paul~AM Van~Lange.
\newblock Avoiding the social death penalty: Ostracism and cooperation in
  social dilemmas.
\newblock {\em The social outcast: Ostracism, social exclusion, rejection, and
  bullying}, pages 321--332, 2005.

\bibitem{pel06}
M.~Peltom\"aki and M.~Alava.
\newblock Correlations in bipartite collaboration networks.
\newblock {\em Journal of Statistical Mechanics: Theory and Experiment.}, page
  P01010, 2006.

\bibitem{pen12}
Jorge Pe{\~n}a and Yannick Rochat.
\newblock Bipartite graphs as models of population structures in evolutionary
  multiplayer games.
\newblock {\em PloS one}, 7(9):e44514, 2012.

\bibitem{ram04}
J.~J. Ramasco, S.~N. Dorogotsev, and R.~Pastor-Satorras.
\newblock Self-organization of collaboration networks.
\newblock {\em Physical Review E}, 70:036106, 2004.

\bibitem{rob08}
Gilbert Roberts.
\newblock Evolution of direct and indirect reciprocity.
\newblock {\em Proceedings of the Royal Society of London B: Biological
  Sciences}, 275(1631):173--179, 2008.

\bibitem{roc09}
C.~P. Roca, J.~A. Cuesta, and A.~S\'anchez.
\newblock Evolutionary game theory: Temporal and spatial effects beyond
  replicator dynamics.
\newblock {\em Physics of Life Review}, 6:208, 2009.

\bibitem{som08}
Ralf~D Sommerfeld, Hans-Juergen Krambeck, and Manfred Milinski.
\newblock Multiple gossip statements and their effect on reputation and
  trustworthiness.
\newblock {\em Proceedings of the Royal Society of London B: Biological
  Sciences}, 275(1650):2529--2536, 2008.

\bibitem{som07}
Ralf~D Sommerfeld, Hans-J{\"u}rgen Krambeck, Dirk Semmann, and Manfred
  Milinski.
\newblock Gossip as an alternative for direct observation in games of indirect
  reciprocity.
\newblock {\em Proceedings of the national academy of sciences},
  104(44):17435--17440, 2007.

\bibitem{ste04}
Pamela~J Stewart and Andrew Strathern.
\newblock {\em Witchcraft, sorcery, rumors and gossip}, volume~1.
\newblock Cambridge University Press, 2004.

\bibitem{sur11}
S.~Suri and D.~J. Watts.
\newblock Cooperation and contagion in web-based, networked public goods
  experiments.
\newblock {\em PLoS ONE}, 6(3):e16836, 2011.

\bibitem{suz05}
S.~Suzuki and E.~Akiyama.
\newblock Reputation in the evolution of cooperation in sizeable groups.
\newblock {\em Proceedings of the Royal Society of London B}, 272:1373--1377,
  2005.

\bibitem{sza07}
G.~Szab\'o and G.~Fath.
\newblock Evolutionary games on graphs. physics reports.
\newblock {\em Physics Reports}, 446:97, 2007.

\bibitem{szo09b}
Attila Szolnoki and Matja{\v{z}} Perc.
\newblock Emergence of multilevel selection in the prisoner's dilemma game on
  coevolving random networks.
\newblock {\em New Journal of Physics}, 11(9):093033, 2009.

\bibitem{szo09c}
Attila Szolnoki and Matja{\v{z}} Perc.
\newblock Resolving social dilemmas on evolving random networks.
\newblock {\em EPL (Europhysics Letters)}, 86(3):30007, 2009.

\bibitem{szo15}
Attila Szolnoki and Matja{\v{z}} Perc.
\newblock Reentrant phase transitions and defensive alliances in social
  dilemmas with informed strategies.
\newblock {\em EPL (Europhysics Letters)}, 110(3):38003, 2015.

\bibitem{szo09}
Attila Szolnoki, Matja{\v{z}} Perc, and Gy{\"o}rgy Szab{\'o}.
\newblock Topology-independent impact of noise on cooperation in spatial public
  goods games.
\newblock {\em Physical Review E}, 80(5):056109, 2009.

\bibitem{tom07}
M.~Tomassini, L.~Luthi, and E.~Pestelacci.
\newblock Social dilemmas and cooperation in complex networks.
\newblock {\em International Journal of Modern Physics C}, 18:1173, 2007.

\bibitem{tra10}
A.~Traulsen, D.~Semmann, R.~D. Sommerfeld, H.~J. Krambeck, and M.~Milinski.
\newblock Human strategy updating in evolutionary games.
\newblock {\em Proceedings of National Academy of Sciences USA},
  107:2962--2966, 2010.

\bibitem{uch13}
S.~Uchida and T.~Sasaki.
\newblock Effect of assessment error and private information on stern-judging
  in indirect reciprocity.
\newblock {\em Chaos, Solitons \& Fractals}, 56:175--180, 2013.

\bibitem{vil11}
D.~Vilone, A.~S\'anchez, and J.~G\'omez-Garde\~nes.
\newblock Random topologies and the emergence of cooperation: the role of
  short-cuts.
\newblock {\em Journal of Statistical Mechanics: Theory and Experiment}, page
  P04019, 2011.

\bibitem{vil14}
Daniele Vilone, Francesca Giardini, and Mario Paolucci.
\newblock {\em New Frontiers in the Study of Social Phenomena}, chapter Partner
  selection supports reputation-based cooperation in a Public Goods Game.
\newblock Springer, 2016.

\bibitem{wil07a}
Kipling~D Williams.
\newblock Ostracism.
\newblock {\em Psychology}, 58(1):425, 2007.

\bibitem{wu16}
Junhui Wu, Daniel Balliet, and Paul~AM Van~Lange.
\newblock Gossip versus punishment: The efficiency of reputation to promote and
  maintain cooperation.
\newblock {\em Scientific reports}, 6, 2016.

\end{thebibliography}

\end{document}